\documentclass[twocolumn,aps,prl,amsmath,amssymb]{revtex4-2}
\usepackage{graphicx}
\usepackage{dcolumn}
\usepackage{bm}

\usepackage{multirow}

\usepackage{amssymb}
\usepackage{amsmath}
\usepackage{graphicx}
\usepackage{xcolor}
\usepackage[colorlinks,citecolor=blue,linkcolor=red,urlcolor=blue]{hyperref}
\usepackage{braket}
\usepackage{CJK}
\usepackage{indentfirst}
\usepackage{amsmath}
\usepackage{cases}

\begin{document}
\title{Nonequilibrium Critical Dynamics with Emergent Supersymmetry}
\author{Zhi Zeng$^{1}$\footnotemark[1]}
\author{Yin-Kai Yu$^{1}$\footnotemark[1]}
\author{Zi-Xiang Li$^{2,3}$\footnotemark[1]}
\email{zixiangli@iphy.ac.cn}
\author{Shuai Yin$^{1}$}
\email{yinsh6@mail.sysu.edu.cn}

\affiliation{$^1$Guangdong Provincial Key Laboratory of Magnetoelectric Physics and Devices, Sun Yat-Sen University, Guangzhou 510275, China}
\affiliation{$^2$Beijing National Laboratory for Condensed Matter Physics \& Institute of Physics, Chinese Academy of Sciences, Beijing 100190, China}
\affiliation{$^3$University of Chinese Academy of Sciences, Beijing 100049, China}

\date{\today}
\begin{abstract}
Proposed as an elegant symmetry relating bosons and fermions, spacetime supersymmetry (SUSY) has been actively pursued in both particle physics and emergent phenomena in quantum critical points (QCP) of topological quantum materials. However, how SUSY casts the light on nonequilibrium dynamics remains open. In this letter, we investigate the Kibble-Zurek dynamics across a QCP with emergent $\mathcal{N}=2$ spacetime SUSY between the Dirac semimetal and a superconductor through large-scale quantum Monte Carlo simulation. The scaling behaviors in the whole driven process are uncovered to satisfy the full finite-time scaling (FTS) forms. More crucially, we demonstrate that the emergent SUSY manifests in the intimate relation between the FTS behaviors of fermionic and bosonic observables, namely the fermions and bosons acquire the identical anomalous dimensions. Our work not only brings a fundamental new ingredient into the critical theory with SUSY, but also provide the theoretical guidance to experimental detect of QCP with emergent SUSY from the perspectives of Kibble-Zurek mechanism and FTS.
\end{abstract}

\maketitle

{\it Introduction}--- Supersymmetry (SUSY), as a symmetry that interchanges bosonic and fermionic fields with each other, was proposed to ingeniously solve the hierarchy problem by exactly canceling the fermion and boson loop corrections to the mass of the Higgs particle~\cite{Weinbergbook,Wessbook,GERVAIS1971632, WESS197439, DIMOPOULOS1981150,NILLES19841,HABER198575}. Although the partner particles predicted by SUSY are still awaiting experimental verification, the elegance of SUSY theory has extensive impacts on modern physics from high-energy physics to condensed matter physics~\cite{PhysRevLett.52.1575,FRIEDAN198537,Zamolodchikov1986,Balents1998NodalLT, PhysRevLett.90.120402, Lee2007EmergenceOS,  Ponte_2014,doi:10.1126/science.1248253,PhysRevLett.114.237001, PhysRevLett.118.166802, PhysRevLett.119.107202, PhysRevLett.115.166401, PhysRevLett.114.090404, PhysRevB.87.165145, PhysRevLett.105.150605, lzx.18sciadv, 
PhysRevLett.126.206801,Franz2019PRB,PhysRevB.100.075153, Gu2024, PhysRevLett.122.240603, Kaviraj2021, PhysRevE.104.044120,PhysRevB.100.075153}. For example, the invariance under local SUSY transformations can automatically reproduce Einstein’s general relativity, resulting in the theory of supergravity~\cite{Wessbook,NILLES19841}. Besides, SUSY and its spontaneous breaking were proposed in both disorder and chaotic systems~\cite{PhysRevE.104.044120,PhysRevLett.122.240603,Kaviraj2021,Junkerbook,Efetovbook}. Moreover, recently, it was theoretically demonstrated that spacetime SUSY spontaneously emerge at some quantum critical points (QCPs) in the quantum many-body systems~\cite{PhysRevLett.52.1575,FRIEDAN198537,Zamolodchikov1986,Balents1998NodalLT, PhysRevLett.90.120402, Lee2007EmergenceOS,  Ponte_2014,doi:10.1126/science.1248253,PhysRevLett.114.237001, PhysRevLett.118.166802, PhysRevLett.119.107202, PhysRevLett.115.166401, PhysRevLett.105.150605,  PhysRevLett.114.090404, PhysRevB.87.165145,  lzx.18sciadv, 
PhysRevLett.126.206801,Franz2019PRB,PhysRevB.100.075153, Gu2024}, for instance (2+1)D spacetime SUSY is proposed to emerge at the superconducting QCP in the boundary of topological insulator hosting single Dirac fermion~\cite{Lee2007EmergenceOS,Ponte_2014,doi:10.1126/science.1248253,lzx.18sciadv}, paving a novel route to experimentally realizing emergent SUSY.

Up to now, SUSY is generally investigated as the equilibrium property of the ground states. However, it is still unclear how SUSY affects the nonequilibrium dynamics, wherein both the ground state and excited states are involved
~\cite{Polkovnikov2011rmp,Dziarmaga2010review,Rigol2016review,Mitra2018arcmp}. This question is motivated by the fact that
nonequilibrium dynamics is ubiquitous in nature from the inflation of the universe in cosmoscopic scale to the collisions of particles in Large Hadron Collider in subatomic scale. In particular, near the critical point, nonequilibrium processes show remarkable universal time-dependent scaling behaviors dictated by the divergent correlation time scale~\cite{Hohenberg1977rmp,Polkovnikov2011rmp,Dziarmaga2010review}. In recent years, programmable quantum devices develop as a practical and tunable platforms to realize QCP~  \cite{king2023nature,Keesling2019,garcia2024resolving,Gu2024}, in which nonequilibrium dynamics is naturally present and utilized to detect quantum critical properties. Hence, unraveling nonequilibrium dynamics of the QCP with emergent SUSY will definitely enrich the theory of supersymmetric quantum criticality and shed new light on the experimental observation of emergent spacetime SUSY at QCP.

Among various nonequilibrium realizations, the celebrated Kibble-Zurek mechanism (KZM) attracts special attentions~\cite{Kibble1976,Zurek1985}. Providing a unified description on the generation and dynamic scaling of topological defects after the linear quench across a critical point~\cite{Kibble1976,Zurek1985}, the KZM has been extensively investigated in cosmological phase transition, classical and quantum phase transitions, from both theoretical and experimental sides~\cite{Kibble1976,Zurek1985,Zurek1997prl,Rajantie2000prl,Chuang1991science,Dziarmaga1998prl,Zoller2005prl,Dziarmaga2005prl,Zurek2007prl,Lamporesi2013,Navon2015science,Du2023,Ko2019,PhysRevLett.125.216601,PhysRevLett.125.100601,PhysRevLett.124.240602,PhysRevLett.127.115701,PhysRevLett.129.227001,PhysRevLett.129.260407,PhysRevLett.131.106501,PhysRevLett.131.230401,PhysRevLett.130.060402,Lamporesi2013_2,Barends2016,King2022,Yao2022,sciadv.aba7292,science.abq6753}. Moreover, dynamic critical behaviors for other quantities are found to exist in the whole driven process~\cite{Zhifangxu2005prb,Deng2008epl,Polkovnikov2010prb,Chandran2012prb,Clark2016science,Huse2012prl}. A finite-time scaling (FTS) theory was proposed with complete scaling forms to understand these scaling properties~\cite{Gong2010njp,huangyy2014prb,Feng2016prb}. Besides, the FTS forms have also been verified in experiments and numerical simulations for different driving protocols~\cite{huangyy2014prb,Feng2016prb,Liuchengwei2014prb,zeng2024FTSGNY,Yin2014prb,Sandvik2015prl,Yin2017prl,Zhong2019pre,Shu2023,Gong2010njp,king2023nature,Keesling2019,Ebadi2021,garcia2024resolving,king2024computational}. Recently, KZM and FTS also show their power in state preparations and the detection of critical properties in fast-developing programmable quantum devices~\cite{king2023nature,Keesling2019,Ebadi2021,garcia2024resolving,king2024computational,PhysRevB.106.L041109,PRXQuantum,science.abo6587}, which provides a promising avenue to experimental realization and detection of QCP featuring exotic critical properties. Given the fundamental importance of SUSY, it is highly desired to investigate driven critical dynamics in the framework of KZM and FTS for the QCP with emergent SUSY.

\begin{figure}[tbp]
\centering
  \includegraphics[width=\linewidth,clip]{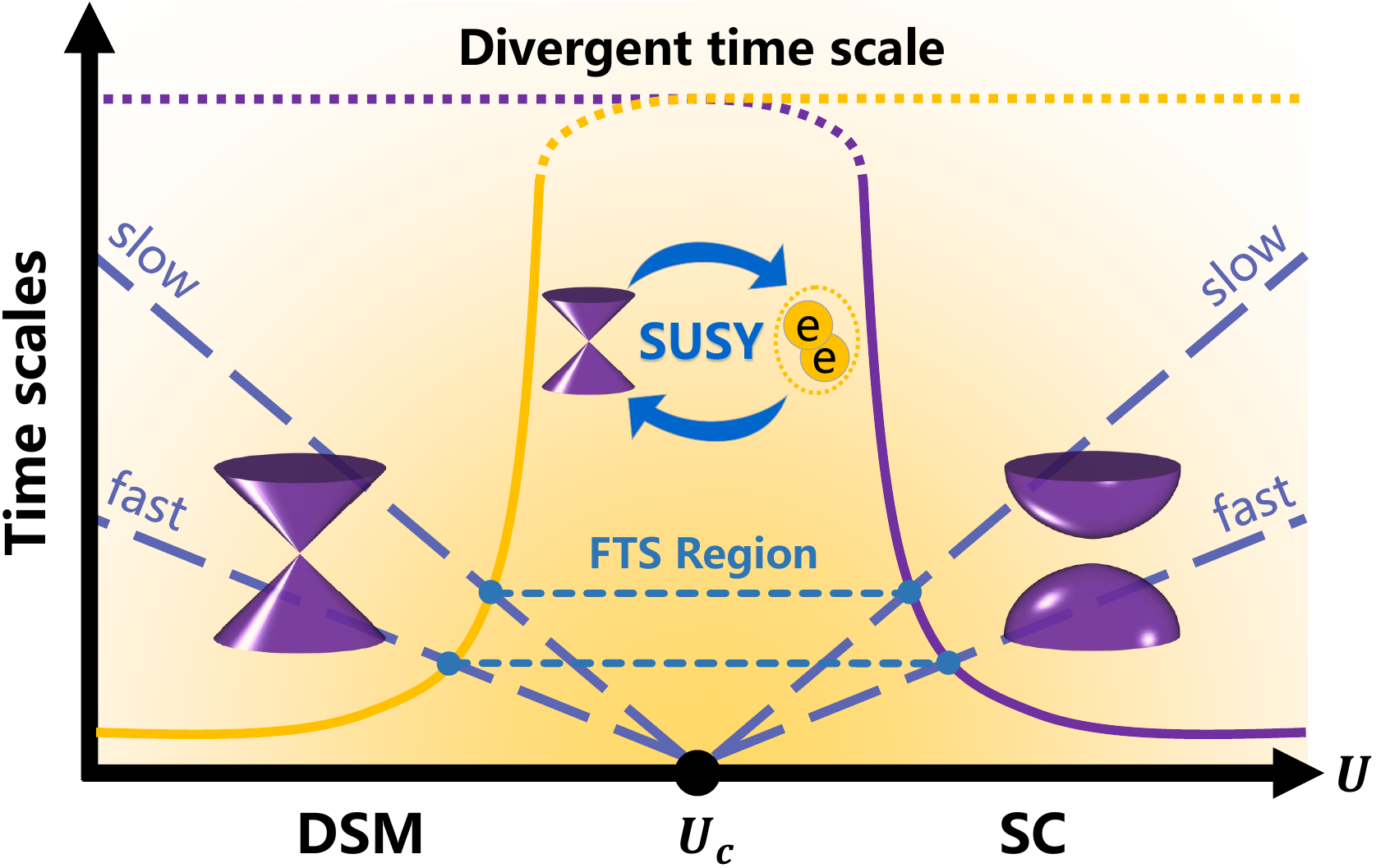}
  \vskip-3mm
  \caption{\textbf{Sketch of the phase diagram with SUSY critical point and the protocol for driven dynamics with different initial states.} The correlation time scales for both boson (yellow curve) and fermion (violet curve) are finite in one phase (solid) but divergent (dotted) in the other phase. Far from the critical point, different from usual KZM, the transition time (long-dashed line) here is smaller than the correlation time of part freedom, but larger than that of the other freedom. Around the critical point, FTS region with emergent SUSY (short-dashed line) dominates the scaling behaviors.
  }
  \label{figure1}
\end{figure}

In this paper, for the first time, we explore the nonequilibrium critical dynamics of a QCP with emergent $\mathcal{N}=2$ SUSY separating a Dirac semimetal (DSM) and a superconducting (SC) phase in a system with SLAC fermions on a square lattice at half filling\cite{SLACfermion,lzx.18sciadv,Lang2019PRL,Abolhassan2022PRL,Assaad2023PRB}, as shown in Fig.~\ref{figure1}. By simulating the driven dynamics under linearly varying interaction strength in imaginary-time direction via the large-scale determinant quantum Monte Carlo (QMC) method~\cite{AssaadReview,lzxqmc_overview},  we uncover that dynamic scaling behaviors depending on the driving rate satisfy the scaling relation of FTS with the critical exponents dictated by SUSY, owing to the principle that both real- and imaginary-time driven dynamics share the same scaling form characterized by the same exponents~\cite{Polkovnikov2011prb,De_Grandi_2013,PolkovnikovSandvik2013prb} which has been verified in various systems~\cite{Polkovnikov2011prb,De_Grandi_2013,PolkovnikovSandvik2013prb,Shu2023,Dziarmaga2022sciadv}. Moreover, we unveil a new dynamic scaling relation of the fermion correlation, which combines the information from both gapless DSM phase and SUSY critical point, generalizing the theory of KZM. Our study fundamentally extends the theory of emergent SUSY to nonequilibrium dynamics which could be realized in programmable quantum processors in the near future~\cite{Gu2024}.

{\it Model and its static criticality}--- We begin with the model which hosts a QCP with emergent SUSY. The Hamiltonian reads~\cite{lzx.18sciadv}
\begin{equation}
 H=\sum_{ij}(t_{\bm{R}}c_{i\uparrow}^\dagger c_{j\downarrow}+H.c.)-U\sum_i \left({n_{i\uparrow}-\frac{1}{2}}\right) \left({n_{i\downarrow}-\frac{1}{2}}\right) \label{eq:Hamiltoniansusy},
\end{equation}
in which $c_{i\sigma}^\dagger$ ($ c_{j\sigma}$) creates (annihilates) an electron at site $\bm{r}_i$ with spin $\sigma=\uparrow/\downarrow$, $n_{i\sigma}\equiv c_{i\sigma}^\dagger c_{i\sigma}$ is the electron number operator, $t_{\bm{R}}(\bm{R}=\bm{r}_i-\bm{r}_j)$ is the amplitude of long-range hopping given by $t_{\bm{R}}= \frac{i(-1)^{\bm{R}_x}}{\frac{L}{\pi}\sin{\frac{\pi \bm{R}_x}{L}}} \delta_{\bm{R}_y,0} + \frac{(-1)^{\bm{R}_y}}{\frac{L}{\pi}\sin{\frac{\pi \bm{R}_y}{L}}} \delta_{\bm{R}_x,0}$, and $U>0$ measures the strength of attractive Hubbard interaction. As shown in Fig.~\ref{figure1}, when $U$ is small, the system is in the DSM phase with a single Dirac point at $\textbf{p}=0$ (namely $\Gamma$ point). Note that this model does not contradict with the fermion-doubling theorem because the hopping here is nonlocal and decaying as $1/r$. In contrast, when $U$ is large, the ground state is the SC state with singlet pairing. In between two phases, there is a QCP at $U = U_c \approx 0.83$ (in unit of the bandwidth)~\cite{lzx.18sciadv}. 

Although the microscopic Hamiltonian~(\ref{eq:Hamiltoniansusy}) does not have SUSY, it was shown that an emergent $\mathcal{N}=2$ SUSY appears at the critical point $U_c$~\cite{lzx.18sciadv}. The most remarkable feature is that anomalous dimensions for both bosonic and fermionic fields obtained via QMC simulations are close to their exact values: $\eta_f = \eta_b = \frac{1}{3}$, where $\eta_f$ and $\eta_b$ are fermion and boson fields anomalous dimension, respectively~\cite{Seiberg1997npb}. This equivalence of fermion and boson anomalous dimensions is a hallmark of SUSY. Moreover, the correlation length exponent $\nu=0.87(5)$ is also consistent with the critical field theory with SUSY~\cite{PhysRevLett.115.051601,PhysRevB.94.205106,PhysRevD.96.096010}. In the following, we will explore the signature of SUSY in the KZM.

{\it Dynamic protocol}--- The quantum KZM focuses on the real-time driven dynamics across a QCP~\cite{Dziarmaga2005prl,Zurek2007prl,Zoller2005prl,Dziarmaga2010review,Polkovnikov2011rmp}. However, simulating the real-time dynamics in higher dimensional systems is still extremely difficult. Fortunately, it was shown that the imaginary-time driven critical dynamics shares the same scaling forms and critical exponents with the real-time case, except for the detailed scaling functions~\cite{Polkovnikov2011prb,De_Grandi_2013,PolkovnikovSandvik2013prb}. The reason is that for both cases, when the driving rate is small, the nonequilibrium dynamics of the system is controlled by the low-lying energy excited states which hold critical properties dominated by the QCP. Accordingly, as the only tuning parameter characterizing the extent of departure from the equilibrium state, the driving rate provides a natural characteristic quantity to describe the nonequilibrium dynamic scaling behaviors in both real-time and imaginary-time directions~\cite{Polkovnikov2011prb,De_Grandi_2013,PolkovnikovSandvik2013prb}. Scaling analyses shows that for both real- and imaginary-time driven dynamics, the driving rate has the same critical dimension. Consequently, one can make a detour to detect the universal scaling properties in the real-time driven process from the imaginary-time dynamics, which can be simulated via the QMC method~\cite{Polkovnikov2011prb,De_Grandi_2013,PolkovnikovSandvik2013prb,Sandvik2015prl,zeng2024FTSGNY}. In the following, we use determinant QMC to simulate the dynamics of model~(\ref{eq:Hamiltoniansusy}) obeying the imaginary-time Schr\"{o}dinger equation $-\frac{\partial}{\partial \tau} |\psi(\tau)\rangle=H(\tau)|\psi(\tau)\rangle$,

in which $g$, defined as $g\equiv U-U_c$, varies linearly with the imaginary time $\tau$ as $g=R \tau$ with $R$ being the driving rate to cross the critical point $U_c$.

\begin{figure*}[ht]
\centering
\includegraphics[width=\textwidth]{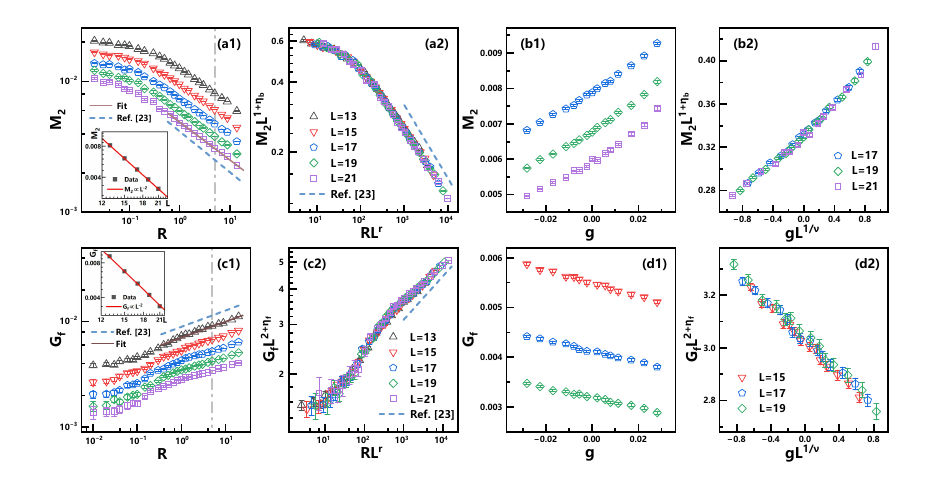}
\caption{\textbf{Driven dynamics from the DSM phase.} \textbf{(a)} Log-log plots of $M_2$ versus $R$ driven to $U_c$ before (a1) and after (a2) rescaling. Inset in (a1) shows $M_2 \propto L^{-2}$ at $R=5$ (dash-dotted line). For large $R$, power fitting for $L=21$ (brown solid line) shows $M_2 \propto R^{-0.285(3)}$ with the exponent close to $(1+\eta_{\mathrm{b}}-d)/r=-0.312$ (dash line) from Ref.~\cite{lzx.18sciadv}. \textbf{(b)} Curves of $M_2$ versus $g$ for fixed $RL^r=347.5$ and different $L$ before (b1) and after (b2) rescaling. \textbf{(c)} Log-log plots of $G_{\mathrm{f}}$ versus $R$ driven to $U_c$ before (c1) and after (c2) rescaling. Inset in (c1) shows $G_{\mathrm{f}} \propto L^{-2}$ at $R=5$ (dash-dotted line). For large $R$, power fitting for $L=13$ (brown solid line) shows $G_{\mathrm{f}} \propto R^{0.152(4)}$ with the exponent close to $\eta_{\mathrm{f}}/r=0.154$ (dash line) from Ref.~\cite{lzx.18sciadv}. \textbf{(d)} Curves of $G_{\mathrm{f}}$ versus $g$ for fixed $RL^r=347.5$ and different $L$ before (d1) and (d2) after rescaling.}
\label{fig:DSM}
\end{figure*}

Moreover, although the original KZM focuses on scaling of topological defects generated after the quench ~\cite{Kibble1976,Zurek1985,Dziarmaga2010review,Polkovnikov2011rmp}, more general scaling behaviors exist for other quantities in the whole driven process~\cite{Zhifangxu2005prb,Deng2008epl,Polkovnikov2010prb,Chandran2012prb,Clark2016science,Huse2012prl}. By identifying an FTS region controlled by the driving-induced time scale $\zeta_R\propto R^{-z/r}$, in which $z$ is the dynamic exponent and $r=z+1/\nu$ with $\nu$ being the correlation length exponent, the FTS theory provides the full scaling forms characterizing the nonequilibrium dynamics near the critical point~\cite{Gong2010njp,Feng2016prb,huangyy2014prb}, as shown in Fig.~\ref{figure1}. Recently, it has been shown that FTS forms are still applicable in Dirac systems with short-range hopping~\cite{zeng2024FTSGNY}. Here we explore the driven dynamics in the critical point with SUSY for SLAC fermions with long-range hopping.

{\it Dynamics with DSM initial state}--- First we study the driven dynamics by increasing $U$ linearly starting from the DSM initial state, as illustrated in Fig.~\ref{figure1}. We begin with the dynamics of the square of the SC order parameter, defined as $M_2\equiv \frac{1}{L^4}\sum_{ij} \braket{\Delta_i^\dagger \Delta_j}$ with $\Delta_i=c_{i\downarrow} c_{i\uparrow}$ the onsite singlet pairing~\cite{lzx.18sciadv}. In the DSM state, it is straightforward to show that $M_2\propto L^{-d}$. At QCP, $M_2$ obeys the scaling $M_2\propto L^{-1-\eta_{b}}$ at equilibrium. For large $R$, this initial state property can be reflected at $U_c$, dictating that $M_2$ obeys according to the scaling analysis:
\begin{equation}
     M_2\propto L^{-d}R^{\frac{1+\eta_{\mathrm{b}}-d}{r}}
    \label{eq:m21}.
\end{equation}
Fig.~\ref{fig:DSM} (a) shows the dependence of $M_2$ on $R$ for different $L$ at $U_c$. Power fitting demonstrates that for large $R$, $M_2$ obeys a power function on $R$ as $M_2\propto R^{-0.285(3)}$, in which the exponent is close to $\frac{1+\eta_{\mathrm{b}}-d}{r}$ with $\eta_{\mathrm{b}}=\frac{1}{3}$, $\nu=0.87$, and $d=2$ set as input. Moreover, with a fixed large $R$, Fig.~\ref{fig:DSM} (a1) shows that $M_2\propto L^{-2}$. These results confirm Eq.~(\ref{eq:m21}). 

Remarkably, the critical exponent of $R$ in Eq.~(\ref{eq:m21}) is a composite of the equilibrium exponents featuring SUSY property, demonstrating that the SUSY can manifest itself via the scaling relation of KZM. In addition, for small $R$, $M_2$ almost saturates to its equilibrium value independent of $R$, satisfying $M_2\propto L^{-1-\eta_{\mathrm{b}}}$~\cite{lzx.18sciadv}. Combing the cases for large $R$ and small $R$, the FTS form for $M_2$ at $U_c$ should be $M_2=L^{-1-\eta_{\mathrm{b}}}\mathcal{F}(RL^r)$, which is verified in Fig.~\ref{fig:DSM} (a2) by substituting $\eta_b=\frac{1}{3}$ and $\nu=0.87$. For large $R$, $\mathcal{F}(RL^r)\propto(RL^r)^{\frac{1+\eta_{\mathrm{b}}-d}{r}}$, as shown in Fig.~\ref{fig:DSM} (a2), presenting the dynamic information of SUSY again. In addition, Eq.~(\ref{eq:m21}) is similar to the KZM scaling relation in the pure boson model~\cite{Liuchengwei2014prb,huangyy2014prb}, demonstrating the universality of KZM critical scaling behavior, regardless of the presence of gapless Dirac fermion.

\begin{figure*}[ht]
\centering
\includegraphics[width=\textwidth]{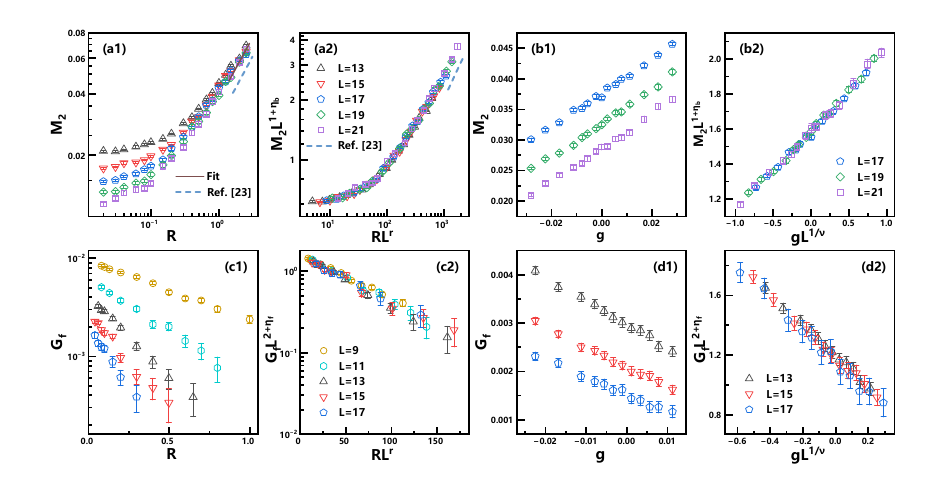}
\caption{\textbf{Driven dynamics from the SC phase.} \textbf{(a)} Log-log plots of $M_2$ versus $R$ driven to $U_c$ before (a1) and after (a2) rescaling. For large $R$, power fitting for $L=21$ (brown solid line) shows $M_2 \propto R^{0.61(3)}$ with the exponent close to $(1+\eta_{\mathrm{b}})/r=0.62$ (dash line) from Ref.~\cite{lzx.18sciadv}. \textbf{(b)} Curves of $M_2$ versus $g$ for fixed $RL^r=347.5$ and different $L$ before (b1) and (b2) after rescaling. \textbf{(c)} Semi-log plots of $G_{\mathrm{f}}$ versus $R$ driven to $U_c$ before (c1) and after (c2) rescaling, where the appearance of straight lines indicate the presence of exponential decay. \textbf{(d)} Curves of $G_{\mathrm{f}}$ versus $g$ for fixed $RL^r=20.85$ and different $L$ before (d1) and (d2) after rescaling.}
\label{fig:SC}
\end{figure*}

Moreover, we show that critical properties with SUSY can be reflected via the FTS in the driven process. In this case, the full scaling form satisfies 
\begin{equation}
M_2(R,L,g)=L^{-d} R^{(1+\eta_{\mathrm{b}}-d)/r} \mathcal{F}_1(RL^{r},gL^{1/\nu}),
\label{eq:m22}
\end{equation}
in which $gL^{1/\nu}$ is included to take account of the off-critical-point effects. By substituting the critical exponents with SUSY, we verify Eq.~(\ref{eq:m22}) in Fig.~\ref{fig:DSM} (b1)-(b2). We note that Eq.~(\ref{eq:m22}) is consistent with the FTS in conventional bosonic QCP~\cite{huangyy2014prb,Liuchengwei2014prb}, and thus demonstrates its universality. 

To further unravel the dynamic scaling with emergent SUSY, it is also instructive to explore the correlation function of fermion operator, which serves as the SUSY partner of $\Delta_i$. The equilibrium finite-size scaling of the fermion correlation $G_{\mathrm{f}}(L)$, defined as $G_{\mathrm{f}}(L)\equiv \frac{1}{L^2}\sum_i\braket{c_i^\dagger c_{i+r_m}+H.c.}$ with $r_m=(\frac{L-1}{2},\frac{L-1}{2})$, is $G_{\mathrm{f}}(L)\propto L^{-d-z+1-\eta_{\mathrm{f}}}$, wherein $\eta_{\mathrm{f}}=\frac{1}{3}$ dictated by SUSY. When driven from the gapless Dirac phase with a large $R$, it is expected that the initial information $G_{\mathrm{f}}(L)\propto L^{-d-z+1}$ of DSM phase should still be remembered at the critical point $U_c$. Accordingly, for large $R$, the scaling relation between $G_{\mathrm{f}}(L)$ and $R$ at $U_c$ should be
\begin{equation}
G_{\mathrm{f}}(R,L)=L^{-d-z+1} R^{\frac{\eta_{\mathrm{f}}}{r}}.
\label{eq:gf1}
\end{equation}

Fig.~\ref{fig:DSM} (c) shows the numerical results of $G_{\mathrm{f}}(R,L)$. We find that for large $R$, $G_{\mathrm{f}}(R,L)$ is a power function of $R$ with the exponent almost independent of $L$. Power fitting shows that the exponent is $0.152(4)$, which is consistent with the results $\frac{\eta_{\mathrm{f}}}{r}=0.154$ within errorbar, given by the critical exponents for $\mathcal{N}=2$ SUSY in 2+1 dimensions\cite{lzx.18sciadv}. Moreover, for fixed $R$, Fig.~\ref{fig:DSM} (c1) shows that $G_{\mathrm{f}}(R,L)\propto L^{-2}$. These results confirm Eq.~(\ref{eq:gf1}). Similar to Eq.~(\ref{eq:m21}), Eq.~(\ref{eq:gf1}) demonstrated that the SUSY can be delivered in the scaling of $R$. Remarkably, it is for the first time to obtain the scaling relation of fermion correlator for the driven dynamics with the DSM initial state, as dictated in Eq.~(\ref{eq:gf1}), which can be generalized to other Dirac fermionic QCPs.

In addition, combining the scaling relation with both large and small $R$, one can obtain the FTS form $G_{\mathrm{f}}=L^{-d-z+1-\eta_{\mathrm{f}}}\mathcal{G}(RL^r)$ in which $\mathcal{G}(RL^r)\propto(RL^r)^{\frac{\eta_{\mathrm{f}}}{r}}$ for large $R$. These results are confirmed in Fig.~\ref{fig:DSM} (c2). Moreover, similar to Eq.~(\ref{eq:m22}), we show that although SUSY emerges exactly at $U_c$, it can affect the dynamic scaling of $G_{\mathrm{f}}$ in the driven process. With $gL^{1/\nu}$ included, the FTS form can be generalized as
\begin{equation}
G_{\mathrm{f}}(R,L,g)=L^{-d-z+1} R^{\eta_{\mathrm{f}}/r} \mathcal{G}_1(RL^{r},gL^{1/\nu}).
\label{eq:gf2}
\end{equation}
For an arbitrarily fixed $RL^{r}$, we verify Eq.~(\ref{eq:m22}) in Fig.~\ref{fig:DSM} (d1)-(d2) by substituting the critical exponents with SUSY.

{\it Dynamics with SC initial state}--- We then explore the dynamic scaling with emergent SUSY from the SC initial state. In Fig.~\ref{fig:SC} (a1), we find that for large $R$, $M_2$ increases with $R$ as $M_2\propto R^{0.61(3)}$ with the exponent close to $\frac{1+\eta_{\mathrm{b}}}{r}$ and almost does not depend on $L$, demonstrating that 
\begin{equation}
     M_2\propto R^{\frac{1+\eta_{\mathrm{b}}}{r}}
    \label{eq:m23}.
\end{equation}
Here the anomalous dimension $\eta_{\mathrm{b}}=\frac{1}{3}$ again enters the scaling relation depending on $R$ as a signature of SUSY. Combining these results with the finite-size scaling for small $R$, one obtains the FTS form at $U_c$ as $M_2=R^{\frac{1+\eta_{\mathrm{b}}}{r}}\mathcal{F}_2(RL^r)$, which is verified in Fig.~\ref{fig:SC} (a2).

By including $gL^{1/\nu}$, the scaling form in the driven process can be generalized as 
\begin{equation}
M_2(R,L,g)=R^{\frac{1+\eta_{\mathrm{b}}}{r}} \mathcal{F}_3(RL^{r},gL^{1/\nu}),
\label{eq:m24}
\end{equation}
which is verified in Fig.~\ref{fig:SC} (b1)-(b2), demonstrating that the scaling with SUSY can appear in the driven process near the critical point.

Then, we explore the dynamic scaling of the fermion correlation $G_{\mathrm{f}}(L)$. For large $R$, we find that $G_{\mathrm{f}}$ at $U_c$ decays as an exponentially decaying function of $R$ as shown in Fig.~\ref{fig:SC} (c1). This is because for larger $R$, more information in the gapped SC phase can be brought to $U_c$. By rescaling $G_{\mathrm{f}}$ and $R$ by $L$ according to the critical exponents with SUSY, we find that the rescaled curves collapse onto each other, as shown in Fig.~\ref{fig:SC} (c2), confirming the FTS form of $G_{\mathrm{f}}(L)$ at $U_c$ is $G_{\mathrm{f}}=L^{-2-\eta_{\mathrm{f}}}\mathcal{G}_3(RL^r)$. In addition, Fig.~\ref{fig:SC} (c2) also shows that the scaling function $\mathcal{G}_3(RL^r)$ satisfies $\mathcal{G}_3(RL^r)\propto \exp(-RL^r)$.

Moreover, we show that SUSY can affect the dynamic scaling of $G_{\mathrm{f}}$ in the driven process. With $gL^{1/\nu}$ included, the FTS form can be generalized as
\begin{equation}
G_{\mathrm{f}}(R,L,g)=L^{-2-\eta_{\mathrm{f}}} \mathcal{G}_4(RL^{r},gL^{1/\nu}).
\label{eq:gf2}
\end{equation}
For an arbitrarily fixed $RL^{r}$, we verify Eq.~(\ref{eq:m22}) in Fig.~\ref{fig:SC} (d1)-(d2) by substituting the critical exponents with SUSY.

{\it The hallmark of emergent SUSY in nonequilibrium dynamics}--- The hallmark of emergent SUSY is the equivalence of anomalous dimensions for boson and fermion. In contrast to previous sections, in which critical exponents are set as input to verify the scaling theory, here we show that the dynamic scaling theory for large $R$ provides an efficient way to determining critical exponents with high accuracy. According to Eq.~(\ref{eq:m21}) and Eq.~(\ref{eq:m23}) and the results  of $M_2$ versus $R$ shown in Fig.~\ref{fig:DSM}(a1) and Fig.~\ref{fig:SC}(a1), we access boson anomalous dimension $\eta_b = 0.36(4)$ and $r = 2.23(8)$. Similarly, from Eq.~(\ref{eq:gf1}) and result in Fig.~\ref{fig:DSM}(c1), the fermion anomalous dimension is achieved $\eta_f = 0.337(14)$. The values of boson and fermion anomalous dimension are identical to each other and consistent with exact values $\eta_b = \eta_f = \frac{1}{3}$ within errorbar, providing convincing evidence of emergent SUSY. Hence, our results unambiguously demonstrate that the nonequilibrium dynamic scaling is a powerful approach to reveal the feature of emergent SUSY, offering a practical tool to detect emergent SUSY in experimental platform.

{\it Summary}--- In summary, we study the driven dynamics of a QCP with emergent SUSY, through sign-problem-free QMC simulation. Driving the system from both DSM and SC phases, we discover interesting nonequilibrium scaling behaviors. In particular, we unveil that SUSY can manifest itself in the scaling relations of KZM depending on the driving rate $R$. Moreover, we show that SUSY also play roles in the driving process by obeying the FTS forms with the critical exponents of SUSY critical point. The scaling relation of the fermion correlation on the driving rate is obtained for the first time, which can be generalized to other Dirac fermionic QCPs. 

Recently, programmable quantum processors have been developed as advanced platforms to realize different phases. In these experiments, the KZM and the FTS are generally used in state preparation. The Kibble-Zurek dynamics and FTS for the QCP with emergent SUSY revealed here provide a practical theoretical framework to detect emergent SUSY in experimental platform. Noting the new progress on the experimental proposal for $(1+1)$D SUSY~\cite{Gu2024}, it is expected that our present results are detectable and potentially helpful for investigating QCP with emergent SUSY in these systems. In addition, our results unambiguously demonstrate that SLAC fermions with long-range hopping still satisfy the nonequilibrium scaling of KZM, although it was shown that special caution should be paid on the unexpected finite-temperature transition related to the excited states~\cite{PhysRevB.108.195112}.

{\it Acknowledgments}--- Z. Zeng, Y. K. Yu, and S. Yin are supported by the National Natural Science Foundation of China (Grants No. 12075324 and No. 12222515). Z. X. Li is supported by the NSFC under Grant No. 12347107. S. Yin is also supported by the Science and Technology Projects in Guangdong Province (Grants No. 2021QN02X561).

%

\onecolumngrid
\newpage
\widetext
\thispagestyle{empty}

\setcounter{equation}{0}
\setcounter{figure}{0}
\setcounter{table}{0}
\renewcommand{\theequation}{S\arabic{equation}}
\renewcommand{\thefigure}{S\arabic{figure}}
\renewcommand{\thetable}{S\arabic{table}}

\pdfbookmark[0]{Supplementary Materials}{SM}
\begin{center}
    \vspace{3em}
    {\Large\textbf{Supplementary Materials for}}\\
    \vspace{1em}
    {\large\textbf{Nonequilibrium Critical Dynamics with Emergent Supersymmetry}}\\
    \vspace{0.5em}
\end{center}

\section{Determinant quantum Monte Carlo}

    Based on the sign-problem free model and that the imaginary-time evolution will not induce additional sign problem, we employ auxiliary field determinant quantum Monte Carlo method~\cite{AssaadReview}, an unbiased numerical algorithm, to study the imaginary-time driven dynamics of our model. 
    
    Starting with an appropriate trial wave function $\ket{\psi_T}$, we can access the accurate ground state of the desired initial state $\ket{\psi(0)}$ through imaginary-time evolution with a sufficiently long time $\tau_0$, which can be written as $|\psi(0)\rangle=\mathrm{e}^{-\tau_0H_0}|\psi_T\rangle$ with $H_0$ being the Hamiltonian of $|\psi(0)\rangle$ corresponding to $U_0$. In our numerical simulation, for DSM initial state we set $U_0=0$, while for SC initial state we set $U_0=1.575$ (in unit of the bandwidth) far from the critical point. For simplicity, we generally choose the ground-state wave function of the given non-interacting Hamiltonian as the trial wave function.
    
    Then we follow $U(\tau)=U_0\pm R \tau$ to vary $U$ and measure physical quantities at final state, where the expectation value of physical quantity $O$ is given by
    \begin{equation}
        \braket{O(\tau)} = 
        \frac{
                \bra{\psi(\tau)}O\ket{\psi(\tau)}
            }{
                \left\langle \psi(\tau) \middle| \psi(\tau) \right\rangle\
            },
    \end{equation}
    where $|\psi(\tau)\rangle = \mathrm{T}~ {\rm exp}\left[-\int_{0}^{\tau}d\tau'H(\tau')\right] |\psi(0)\rangle$ with $\mathrm{T}$ the time-ordering operator in imaginary-time direction. 
    
    We write the Hamiltonian as $H=H_t+H_U$, where $H_t$ is the hopping term and $H_U$ is the interaction term.
    In numerical calculations, we employ Trotter decomposition to discretize imaginary time, which in driven dynamics is written as 
    \begin{equation}
        \mathrm{T}~ {\rm exp}\left[-\int_{0}^{\tau}d\tau'H(\tau')\right] = \lim_{\Delta\tau \to 0} \prod^{M}_{n=1} \left[\mathrm{e}^{-\Delta\tau H_t} \mathrm{e}^{-\Delta\tau H_U(\tau_n)} \right],
    \end{equation}
    in which the imaginary-time discretisation $\Delta\tau=\tau/M=0.05$ and $\tau_n=n\Delta\tau$.

    For the Hubbard interaction term, we convert it into a perfect square and decouple the fermion-fermion coupling term to a fermion-auxiliary field coupling term using discrete Hubbard-Stratonovich (HS) transformation:
        
    \begin{equation}
    \begin{split}
        &\mathrm{e}^{\Delta\tau U(\tau_n)(n_{i\uparrow}-\frac{1}{2})(n_{i\downarrow}-\frac{1}{2})}\\
        =~
        &\mathrm{e}^{-\frac{1}{4}\Delta\tau U(\tau_n)}~\mathrm{e}^{\frac{1}{2}\Delta\tau U(\tau_n)\left(n_{i\uparrow}+n_{i\downarrow}-1\right)^2}\\
        =~
        &\frac{1}{2}~\mathrm{e}^{-\frac{1}{4}\Delta\tau U(\tau_n)} \sum_{s=\pm 1}\mathrm{e}^{\alpha s\left(n_{i\uparrow}+n_{i\downarrow}-1\right)},
        \label{seq:HS}
    \end{split}
    \end{equation}
    where $\mathrm{cosh}(\alpha) = \mathrm{e}^{\Delta \tau U /2}$. In Eq.~(\ref{seq:HS}), we choose two-component space-time auxiliary fields considering its faster Monte Carlo equilibration and no systematic error. It is worth mentioning that for negative prefactor ($U>0$), Eq.~(\ref{seq:HS}) involves only real numbers.

    At this stage, the one-body fermions operators can be integrated out and we obtain

    \begin{equation}
        \left\langle \psi(\tau) \middle| \psi(\tau) \right\rangle\ 
        = \sum_{\Vec{s}} C_{\Vec{s}} ~\mathrm{det} \left[P^\dagger B_{\Vec{s}}(2\tau,0)P\right],
        \label{seq:det}
    \end{equation}
    in which $C_{\Vec{s}}$ is the coefficient, $P$ is the initial state with fermionic degrees of freedom intergrated out and
    \begin{equation}
        B_{\Vec{s}}(2\tau,0) \equiv \prod^{2M}_{n=1} ~ \mathrm{e}^{h_I(\Vec{s}_n)} ~ 
        \mathrm{e}^{-\Delta \tau h_t},
    \end{equation}
    with $h_I(\Vec{s}_n)$ and $h_t$ being the quadratic coefficient matrices of interaction term and hopping term respectively.
    
    After the aforementioned operations, we sample over the space-time configurations to obtain the expectation value of observables:
    \begin{equation}
        \braket{O(\tau)} = \sum_{\Vec{s}} \mathrm{Pr}_{\Vec{s}} \braket{O(\tau)}_{\Vec{s}} + \mathcal{O}\left(\Delta\tau^2\right),
    \end{equation}
    where $\braket{O(\tau)}_{\Vec{s}}$ represents the value of observables in certain auxiliary field configuration
    \begin{equation}
        \braket{O(\tau)}_{\Vec{s}} = 
        \frac{
                \bra{\psi_T}\mathrm{e}^{-\tau_0 H_0}\overline{\mathrm{T}}~ {\rm exp}\left[-\int_{0}^{\tau}d\tau'H_{\Vec{s}}(\tau')\right]~O~\mathrm{T}~ {\rm exp}\left[-\int_{0}^{\tau}d\tau'H_{\Vec{s}}(\tau')\right]
                \mathrm{e}^{-\tau_0 H_0}
                \ket{\psi_T}
            }{
                \left\langle \psi(\tau) \middle| \psi(\tau) \right\rangle\
            },
            \end{equation}
    and $\mathrm{Pr}_{\Vec{s}}$ is the configuration probability
    \begin{equation}
        \mathrm{Pr}_{\Vec{s}} = \frac{C_{\Vec{s}} ~\mathrm{det} \left[P^\dagger B_{\Vec{s}}(\tau,0)P\right]}{\sum_{\Vec{s}} C_{\Vec{s}} ~\mathrm{det} \left[P^\dagger B_{\Vec{s}}(2\tau,0)P\right]}.
    \end{equation}

\end{document}